\documentclass[letterpaper]{article}
\usepackage[ansinew]{inputenc}
\usepackage{amsmath}
\usepackage{amsthm}
\usepackage{mathtools}
\usepackage{amssymb}
\usepackage{enumerate}
\usepackage{hyperref}
\usepackage{bbm}
\usepackage[margin=0.8in]{geometry}
\usepackage{setspace}

\usepackage{authblk}

\title{Undecidability and Irreducibility Conditions for Open Ended Evolution and Emergence}

\author[1]{Santiago Hern\'{a}ndez-Orozco\thanks{Contact author: hosant@ciencias.unam.mx}}
\author[2]{Francisco Hern\'{a}ndez-Quiroz\thanks{fhq@ciencias.unam.mx}}
\author[3,4,6]{Hector Zenil\thanks{hector.zenil@algorithmicnaturelab.org}}
\affil[1]{Posgrado en Ciencia e Ingenier\'ia de la Computaci\'on, UNAM, Mexico.}
\affil[2]{Departamento de Matem\'aticas, Facultad de Ciencias, UNAM, Mexico.}
\affil[3]{Department of Computer Science, University of Oxford, UK.}

\affil[4]{Unit of Computational Medicine, SciLifeLab, Department of Medicine Solna\\ Centre for Molecular Medicine, Stockholm, Sweden.}

\affil[5]{Algorithmic Nature Group, LABORES, Paris, France.}

\begin{document}

{
\singlespacing
\maketitle
}

\theoremstyle{plain}
\newtheorem{thm}{Theorem}
\newtheorem{lem}[thm]{Lemma}
\newtheorem{cor}[thm]{Corollary}
\newtheorem{conj}[thm]{Conjecture}
\theoremstyle{definition}
\newtheorem{defn}[thm]{Definition}

\begin{abstract}
Is undecidability a requirement for open-ended evolution (OEE)?
Using methods derived from algorithmic complexity theory, we propose robust
 computational definitions of open-ended evolution and the 
adaptability of computable dynamical systems. Within this
 framework, we show that decidability imposes absolute limits to
 the stable growth of complexity in computable dynamical systems. 
Conversely, systems that exhibit
(strong) open-ended evolution must be undecidable, establishing 
undecidability as a requirement for such systems. Complexity is
assessed in terms of three measures: sophistication, coarse 
sophistication and busy beaver logical depth. These three
complexity measures assign low complexity values to random 
(incompressible) objects. As time grows, the stated complexity measures allow for the existence of complex states during the evolution of a computable dynamical system. We show, however, that finding these states involves undecidable computations. We conjecture that for similar complexity measures that assign low complexity values, 
decidability imposes comparable limits to the stable growth of 
complexity, and that such behaviour is necessary for non-trivial 
evolutionary systems. We show that the undecidability of 
adapted states imposes novel and unpredictable behaviour on the 
individuals or populations being modelled. Such behaviour is 
irreducible. Finally, we offer an example of a system, first proposed by Chaitin, 
that exhibits strong OEE.

\end{abstract}

\section{Introduction and Preliminaries}

Broadly speaking, a dynamical system is one that changes over
time. Prediction of the future behaviour of dynamical systems is 
a fundamental concern of science generally. Scientific theories are tested 
upon the accuracy of their predictions, and establishing 
invariable properties through the evolution of a system is an
 important goal. Limits to this predictability are known in 
science. For instance, chaos theory establishes the existence of 
systems in which small deficits in the information of the initial 
states makes accurate predictions of future states unattainable. 
However, in this document we focus on systems for which we have 
unambiguous, finite (as to size and time) and complete descriptions
 of initial states and behaviour: computable dynamical
 systems.

 Since their formalization by Church and Turing, the class of 
computable systems has shown that, even without information 
deficits (i.e., with complete descriptions), there are future
 states that cannot be predicted, in particular the state known as the
\emph{halting state} \cite{Turing}. We will use this result and 
others from algorithmic information theory to show how
predictability imposes limits to the growth of complexity during 
the evolution of computable systems. In particular, we will show that 
random (incompressible) times tightly bound the complexity of the 
associated states. 

The relationship between dynamical systems and computability has been 
studied before by Bournez \cite{Bournez, CIE2013Daniel}, Blondel \cite{BBKT00},
Moore \cite{Moore} and by Fredkin, Margolus and Toffoli \cite{Fredkin-Toffoli-82,Margolus84},
 among others. That emergence is a consequence of incomputability has been proposed by 
Cooper \cite{Cooper:Emergence}. Complexity as a source of undecidability has been 
observed in logic by Calude and Jurgensten \cite{complexIncomplt}. 
Delvenne, Kurka and Blondel \cite{SymbDynSystm} have proposed robust definitions of
 computable (effective) dynamical systems and universality, 
generalizing Turing's halting states, while also setting forth the conditions and implications for universality and decidability and their relationship 
with chaos. The definitions and general approach used in this 
paper differ from those in the sources cited above, but are ultimately related. 

We will denote by $K(x|y)$ the algorithmic descriptive complexity of the string $x$ with respect to the string $y$. The dynamical systems we are considering are deterministic, and each state must contain all the information needed to compute successive states. We are assuming an infinity of possible states for non-cyclical systems. Mechanisms and requirements for open-ended evolution in systems with a finite number of states (resource-bounded) have been studied by Adams et al. \cite{alyssa}.

\subsection{Computable Functions}

In a broad sense, an object $x$ is \textit{computable} if it can 
be described by a Turing machine \cite{Turing}; for example, if 
there exists a Turing machine that produces $x$ as an output. It is 
clear that any finite string on a finite alphabet is a computable 
object. We provide below a more 
formal definition, in the tradition of Turing.

 As usual, we can define a one-to-one mapping between the set of all 
finite binary strings $\mathbb{B}^*=\{0,1\}^*$ and the natural
 numbers by the relation induced by a lexicographic order of the
 form: $\{(\text{``''},0), (\text{``0''},1), (\text{``1''},2),
(\text{``00''},3),...\}$. Using this relation we can see all 
natural numbers (or positive integers) as binary strings and vice 
versa. Accordingly all natural numbers are computable. 

A string $p$ is a \textit{valid program} for the Turing machine 
$T$ if during the execution of $T$ with $p$ as input all the 
characters in $p$ are read. We call $T(p)$ the output of the
machine, if it stops. A Turing Machine is \textit{prefix-free} if 
no valid program can be a proper substring of another valid
 program (though it can be a postfix of one). We call a valid program a
\textit{self-delimited object}. Note that, given the relationship
 between natural numbers and binary strings, the set of all valid 
programs is an infinite proper subset of the natural numbers.

 Formally, a function $f:\mathbb{N} \rightarrow \mathbb{N}$ is
\emph{computable} if there exists a Turing Machine $T$ such that
$f(x)=T(x)$. A Turing Machine $U$ is considered \textit{universal} if 
there exists a computable function $g$ such that for every Turing 
machine $T$ there exists a string $\langle T \rangle \in
\mathbb{B}^*$ such that $f(x)=U(\langle T \rangle g( x ))$, where
$\langle T \rangle g( x )$ is the concatenation of the strings
$\langle T \rangle$ and $g( x )$. Given the previous case, 
$\langle T \rangle$ and $g(x)$ are called a \textit{codification
 or a representation} of the function $f$ and the natural number
$x$, respectively. From now on we will denote the codification of $f$ and $x$ by $\langle f
\rangle$ and $\langle x \rangle$.
The codification $g(x)$ is \emph{unambiguous} if it is injective. 

For functions with more than one variable, if $x$ is a pair
$x=(x_1,x_2)$, we say that the codification $g(x)$ is unambiguous 
if it is injective and the inverse functions $g^{-1}_1: g(x) \mapsto
x_1$ and $g^{-1}_2: g(x) \mapsto x_2$ are computable. If $x$ is a
tuple $(x_1,...,x_i,...,x_n)$, then the codification $g(x)$ is
unambiguous if the function $(x,i) \mapsto x_i$ is computable.

 A sequence of strings $\delta_1,\delta_2,...,\delta_i,...$ is
 computable if the function $\delta : i \mapsto \delta_i$ is 
computable. A real number is computable if its decimal expansion
is a computable sequence. For complex numbers and higher 
dimensional spaces, we say that they are computable if each of their 
coordinates is also computable. 

Finally, for each of the objects described, we refer to the 
representation of the associated Turing machine as  \emph{the 
representation of the object for the reference Turing machine
$U$}, and we define the computability of further objects by 
considering their representations. For example, a function
$f:\mathbb{R} \rightarrow \mathbb{R}$ is computable if the mapping
$\langle x_i \rangle \mapsto \langle f(x_i) \rangle$ is computable
and we will denote by $\langle f \rangle$ the representation of the 
associated Turing machine, calling it the codification of $f$ 
itself.

\subsection{Algorithmic Descriptive Complexity}

Given a prefix-free universal Turing Machine $U$ with alphabet 
$\Sigma$, the \textit{algorithmic descriptive complexity} (also 
known as Kolmogorov complexity and Kolmogorov-Chaitin complexity
\cite{Kolmogorov,Chaitin3}) of a string $s\in{}\Sigma^*$ is 
defined as
\begin{equation*}
K_U(s) = \min\{|p|:U(p)=s\},
\end{equation*}
where $U$ is a universal prefix-free Turing Machine and $|p|$ is 
the number of characters of $p$. 

Algorithmic descriptive complexity measures the minimum amount
of information needed to fully describe a computable object within 
the framework of a universal Turing machine $U$. If $U(p)=s$ then 
the program $p$ is called a description of $s$. The first of the
 smallest descriptions (in alphabetical order) is denoted by $s^*$ 
and by $\langle{}s\rangle{}$, a not necessarily minimal description
 computable over the class of objects. If $M$ is a Turing machine,
 a program $p$ is a description or codification of $M$ for $U$ if 
for every string $s$ we have it that $M(s) = U(p \langle s \rangle)$.
In the case of numbers, functions, sequences and other computable 
objects we consider the descriptive complexity of their smallest 
descriptions. For example, for a computable function $f:\mathbb{R}
\rightarrow \mathbb{R}$, $K(f)$ is defined as $K(f^*)$, where
$f^*\in \mathbb{B}^*$ is the first of the minimal descriptions for
$f$.

\noindent{}Of particular importance for this document is the
\emph{conditional descriptive complexity}, which is defined as:
\begin{equation*}
K_U(s|r) = \min \{|p|:U(pr)=s\},
\end{equation*}
where $pr$ is the concatenation of $p$ and $r$. This measure can
be interpreted as the \emph{smallest amount of information needed
to describe $s$ given a full description of $r$}. We can think of
$p$ as a program with input $r$. 

One of the most important properties of the descriptive complexity 
measure is its \emph{stability}: the difference between the 
descriptive complexity of an object, given two universal Turing 
machines, is at most constant. Therefore the reference machine $U$
is usually omitted in favor of the \textit{universal measure} $K$.
From now on we will omit the subscript from the measure.

\subsubsection{Randomness}

Given a natural number $r$, a string $x$ is considered 
$r$\emph{-random} or \emph{incompressible} if $K(x) \geq |x| - r$. 
This definition would have it that a string is random if it does not have 
a significantly shorter complete description than the string 
itself. A simple counting argument shows the existence of random 
strings. Now, it is easy to verify that every string $x$ has a self
-delimited, unambiguous computable codification with strings of the 
form $1^{\log |s|}0|s|s$ ($\log |s|$ 1s followed by a 0, then the
 binary string corresponding to $|s|$ concatenated with the string
 $s$ itself \cite[section 1.4]{IntKolmogorov}). Therefore, there 
exists a natural $r$ such that if $x$ is $r$-random then $K(x) =
|x| - r + O(\log |x|)$, where $O(\log |x|)$ is a positive term. We 
will say that such strings hold the randomness inequality
\emph{tightly}.

Let $M$ be a halting Turing Machine with description $\langle M
\rangle$ for the reference machine $U$. A simple argument can show
t that the halting time of $M$ cannot be a large \emph{random}
number. Let $U^H$ be a Turing Machine that emulates $U$ while
 counting the number of steps, returning the execution time upon 
halting. If $r$ is a large random number, then $M$ cannot stop in 
time $r$, otherwise the program $\langle U^H \rangle \langle M
\rangle$ will give us a \emph{short} description of $r$. This argument 
is summarized by the following inequality:
\begin{equation}\label{primera}
K(T(M))\leq K(M) + O(1),
\end{equation}
where $T(M)$ is the number of steps that it took the machine $M$ to
 reach the halting state, the \emph{execution time} of the machine
 $M$.\\

\subsection{Computable Dynamical Systems}

Formally, a \textit{dynamical system} is a rule of evolution in 
time within a state space; a space that is defined as the set of all 
possible states of the system \cite{dynamicS}. In this paper we 
will focus on a functional model for dynamical systems with a 
constant initial state and variables representing the previous 
state and the time of the system. This model allows us to set 
halting states for each time on a discrete scale in order to study 
the impact of the descriptive complexity of time during the
evolution of a discrete computable system. 

A deterministic discrete space system is defined by an
\emph{evolution function (or rule)} of the form $M_{t+1} = S(M_0,
t)$, where $M_0$ is called the \emph{initial state} and $t$ is a
positive integer called the \textit{time} variable of the system.
The sequence of states $M_0,M_1,...,M_t,...$ is called the
\text{evolution} of the system. Given a reference universal Turing 
Machine $U$, if $S$ is a computable function and $M_0$ is a
computable object, we will say that $S$ is a \emph{computable
dynamical system}. An important property of computable dynamical 
systems is the uniqueness of the successor state, which implies 
that equal states must evolve equally given the same evolution 
function. In other words:
\begin{equation}\label{uSucP}
M_t=M_{t'} \implies M_{t'+1}=M_{t+1}.
\end{equation}
\noindent{}The converse is not necessarily true.

Now, a \emph{complete description of a computable system} $S(M_0,
t)$ should contain enough information to compute the state of the 
system at any time and hence it must entail the codification of
its evolution function $S$ and a description of the initial state
$M_0$, which is denoted by $\langle M_0 \rangle$. As a consequence, if 
we only describe the system at time $t$ by a codification of
$M_t$, then we would not have enough information to compute the successive states of the system. So we will specify the
\emph{complete description} of a computable system at time $t$ 
as an unambiguous codification of the ordered pair composed of 
$\langle S \rangle$ and $\langle M_t \rangle$, i.e. $\langle
(S,\langle M_t \rangle) \rangle$, with $\langle (S,\langle M_0
\rangle) \rangle$ representing the initial state of the system. It 
is important to note that, for any computable and unambiguous 
codification function $g$ of the stated pair, we have
$K(\langle (S,\langle M_t \rangle)) \leq K(S) + K(M_0) + K(t) +
O(1)$, as we can write a program that uses the descriptions
for $S$, $M_0$ and $t$ to find the parameters and then evaluate 
$S(M_0, t)$, finally producing $M_t$. 

It is important to mention that, given that the dynamical systems we are considering are deterministic, and that each state must contain all the information needed to compute successive states, we are assuming an infinity of possible states for non-cyclical systems. Mechanisms and requirements for open-ended evolution in systems with a finite number of states (resource-bounded) have been studied by Adams et al. \cite{alyssa}.

\subsection{Open-Ended Evolution in Computable Dynamical Systems}

Informally, \textbf{Open-ended evolution (OEE)} has been 
characterized as ``evolutionary dynamics in which new, surprising,
and sometimes more complex organisms and interactions continue to
appear'' \cite{TaylorOEEW}. Establishing and defining the 
properties required for a system to exhibit OEE is considered an 
open question \cite{BedauEtAl00,Soros:2014,openEndedArtEv} and 
OEE has been proposed as a required property of evolutionary 
systems capable of producing life \cite{Univdeflife}. This has
 been implicitly verified by various experiments \emph{in-silico}
\cite{lindgren:1992:epsd, Adami94a,expltOEE, AuerbachB14}.

 One line of thought posits that open-ended evolutionary systems tend 
to produce families of objects of increasing \emph{complexity}
\cite{Bedau98,AuerbachB14}. Furthermore, for a number of
complexity measures, it can be shown that the objects belonging to 
a given level of complexity are finite (for instance $K(x)$). Therefore 
an increase of complexity is a requirement for the continued production of new
 objects. A related observation, proposed by Chaitin
\cite{chaitin:EvolofMutaSoft,ChaitinBook}, associates evolution 
with the search for \emph{mathematical creativity}, which implies 
an increase of complexity, as more complex mathematical operations 
are needed in order to solve \emph{interesting problems}, which 
are \emph{required to drive evolution}. 

Following the aforementioned lines of thought, we have chosen to characterize OEE in 
computable dynamical systems as a process that has the property of 
producing families of objects of increasing \emph{complexity}.
Formally, given a \textit{complexity measure $C$}, we say that a 
computable dynamical system $S$ exhibits \textit{open-ended 
evolution} with respect to $C$ if for every time $t$ there exists 
a time $t'$ such that the complexity of the system at time
$t'$ is greater than the complexity at time $t$, i.e.
$C(S(M_0,t)) < C(S(M_0,t')$, where a complexity measure is a (not 
necessarily computable) function that goes from the state space to 
a positive numeric space.

 The existence of such systems is trivial for complexity measures 
on which any infinite set of natural numbers (not necessarily 
computable) contains a subset where the measure grows strictly:

\begin{lem}\label{trivialOEE}
Let $C$ be a complexity measure such that any infinite set of 
natural numbers has a subset where $C$ grows strictly. Then a
computable system $S(M_0,t)$ is a system that produces an infinite 
number of different states if and only if it exhibits OEE for $C$.

\begin{proof}

Let $S(M_0,t)$ be a system that does not exhibit OEE, and $C$ a 
complexity measure as described. Then there exists a time $t$ such
 that for any other time $t'$ we have $C(M_t) \leq C(M_{t'})$, 
which holds true for any subset of states of the system. It follows 
that the set of states must be finite. Conversely, if the system exhibits OEE, then there exists an 
infinite subset of states on which $S$ grows strictly, hence an infinity of different states.

\end{proof}
\end{lem}

Given the previous lemma, a trivial computable system that simply
produces all the strings in order exhibits OEE on a class of 
complexity measures that includes algorithmic description
 complexity. However, we intuitively conjecture that such systems 
have a much simpler behaviour compared to that observed in the 
natural world and the artificial life systems referenced. To avoid 
some of these issues we propose a stronger version of OEE.
\begin{defn}
A sequence of naturals $n_0,n_1,...,n_i,...$ exhibits \emph{strong 
open-ended evolution} (strong OEE) with respect to a complexity 
measure $C$ if for every index $i$ there exists an index $i'$ such
that $C(n_i) < C(n_{i'})$, and the sequence of complexities
$C(n_0),C(n_1),...,C(n_i),...$ does not drop \textit{significantly}, i.e. there exists 
a $\gamma$ such that $i \leq j$
implies $C(n_i) \leq C(n_j) + \gamma(j)$ where $\gamma(j)$ is a positive function that 
does not grow \textit{significantly}.
\end{defn}
\noindent{}It is important to note that while the definition of 
OEE allows for significant drops in complexity during the 
evolution of a system, strong OEE requires that the complexity of
 the system not decrease \emph{significantly} during its 
evolution. In particular we will require that the \textit{complexity 
drops} as measured by $\gamma$  not \emph{grow as fast as the 
complexity itself} and that they reach a constant level an infinite number of times. 
Formally $C(n_j)-\gamma(j)$ should not be
upper-bounded for any infinite subsequence for the smallest 
$\gamma$ where the strong OEE inequality holds. 

We will construe the concept of \emph{speed} of growth of 
complexity in a comparative way: given two sequences of natural 
numbers $n_i$ and $m_i$, $n_i$ \emph{grows faster} than $m_i$ if 
for every infinite subsequence and natural number $N$, there exists 
$j$ such that $n_i-m_j \geq N$. Conversely, a subsequence of 
indexes denoted by $i$ grows faster than a subsequence of indexes
 denoted by $j$ if for every natural $N$, there exists $i$ with 
$i<j$, such that $n_i - n_j \geq N$. 

If a complexity measure is sophisticated enough to depend on more 
than just the size of an object, significant drops in complexity 
are a feature that can be observed in trivial sequences such as 
the ones produced by enumeration machines. Whether this is also 
true for \emph{non-trivial} sequences is open to debate. However,
if we classify random strings as low complexity objects and posit
that non-trivial sequences must contain a limited number of random 
objects, then a non-trivial sequence must observe bounded drops in 
complexity in order to be capable of showing non-trivial OEE. This
 is the intuition behind the definition of strong OEE. 

Now, in the literature on dynamical systems, 
random objects are often considered simple (\cite[pp.1]{adami2002complexity}), with complexity being taken to lie between regularity and randomness. Various complexity measures have been proposed that assign
 low complexity to random or incompressible natural numbers. Two 
examples of such measures are logical depth \cite{LDepth} and
sophistication \cite{koppelSoph}. Classifying random naturals as
low complexity objects is a requirement for the results shown in
section \nameref{OOE}.

\section{A Computational Model for Adaptation}

Let's start by describing the evolution of an organism or a 
population by a computable dynamical system. It has been argued
 that in order for \textit{adaptation} and survival to be 
possible an organism must contain an effective representation of the
environment, so that, given a reading of the environment, the 
organism can choose a behaviour accordingly \cite{zenil1}. The 
more approximate this representation, the better the adaptation. If the organism is computable, this
 information can be codified by a computable structure. We will
 denote this structure by $M_t$, where $t$ stands for the time
 corresponding to each of the stages of the evolution of the 
organism. This information is then processed following a finitely
 specified unambiguous set of rules that, in finite time, will 
determine the adapted behaviour of the organism according to the 
information codified by $M_t$. We will denote this behaviour (or a 
theory explaining it) using the program $p_t$. An adapted system is
one that produces an acceptable approximation of its environment. 
An environment can also be represented by a computable structure
$E$. In other words, the system is adapted if $p_t(M_t)$ produces
$E$. Based on this idea we propose a robust, formal characterization for
adaptation:
\begin{defn}\label{adaptation}
Let $K$ be the prefix-free descriptive complexity. We say that the 
system at the state $M_n$ is \emph{$\epsilon$-adapted} to the $E$
if:
\begin{equation}
K(E|S(M_0,E(n))) \leq \epsilon.
\end{equation}

The inequality states that the minimal amount of information that 
is needed to describe $E$ from a complete description of $M_n$ is
$\epsilon$ or less. This information is provided in the form of a 
program $p$ that produces $E$ from the system at time $n$. We
 will define such a program $p$ as the \emph{adapted behaviour} of 
the system. It is not required that $p$ be unique.
\end{defn}

The proposed structure for adapted systems is robust since 
$K(E|S(M_0,E,n))$ is less than or equal to the number of characters 
needed to describe any computable method of describing $E$ from 
the state of the system at time $n$, whether it be a computable
 theory for adaptation or a computable model for an organism that 
tries to predict $E$. It follows that any computable characterization 
of adaptation that can be described within $\epsilon$ number of 
bits meets the definition of \emph{$\epsilon$-adapted}, given
 a suitable choice of $E$, the \emph{adaptation condition} for any 
given environment. It is important to note that, although inspired by 
a representationalist approach to adaptation, the proposed 
characterization of adaptation is not contingent on the organism
's containing an actual codification of the environment, since 
any organism that can produce an adapted behaviour that can be
 explained effectively (is computable in finite time) is 
$\epsilon$-adapted for some $\epsilon$. 

As a simple example, we can think of an organism that must find 
food located at the coordinates $(x,j)$ on a grid in order to 
survive. If the information in an organism is codified by a 
computable structure $M$ (such as DNA), and there is a set of 
finitely specified, unambiguous rules that govern how this
 information is used (such as the ones specified by biochemistry 
and biological theories), codified by a program $p$, then we say 
that the organism finds the food if $p(M)= (j,k)$. If $|\langle p
\rangle| \leq \epsilon$, then the we say that the organism is 
adapted according to a behaviour that can be described within
$\epsilon$ characters. The proposed model for adaptation is not
 limited to such simple interactions. For a start, we can suppose 
that the organism  \textit{sees} a grid, denoted by $g$, of size
$n \times m$ with food at the coordinates $(j,k)$. The environment 
can be codified as a function $E$ such that $E (g) = (j,k)$ and
$\epsilon$-adapted implies that the organism defined by the 
genetic code $M$, which is interpreted by a theory or behaviour 
written on $\epsilon$ bits, is capable of finding the food upon 
seeing $g$. Similarly, more complex computational structures and
interactions imply $\epsilon$-adaptation. 

Now, describing an evolutionary system that (eventually) produces 
an \emph{$\epsilon$-adapted} system is trivial via an enumeration 
machine (the program that produces all the natural numbers in
order), as it will eventually produce $E$ itself. Moreover, we 
require the output of our process to remain adapted. Therefore we 
propose a stronger condition called \emph{convergence}:

\begin{defn}\label{convergence}
Given the description of a computable dynamical system
 $S(M_0,E,t)$ where $t\in{}\mathbb{N}$ is the variable of time, 
$M_0$ is an initial state and $E$ is an environment, we say that 
the system $S$ \emph{converges} towards $E$ with degree $\epsilon$ 
if there exists $\delta$ such that $t \geq \delta$ implies
$K(E|S(M_0,E,t)) \leq \epsilon$.
\end{defn}
For a fixed initial state $M_0$ and environment $E$, it is easy to 
see that the descriptive complexity of a state of the system 
depends mostly on $t$: we can describe a program that, given full 
descriptions of  $S$, $E$, $M_0$ and $t$, finds $S(M_0,E,t)$. Therefore
\begin{equation}\label{timeDomIneq}
K(S(M_0,E,t)) \leq K(S) + K(E) + K(M_0) + K(t) + O(1),
\end{equation}
where the constant term is the length of the program described. In 
other words, as the time $t$ grows, \emph{time becomes the main driver
 for the descriptive complexity within the system}.

\subsection{Irreducibility of Descriptive Time Complexity}

In the previous section, it was established that time was the main 
factor in the descriptive complexity of the states within the
 evolution of a system. This result is expanded by the time 
complexity stability theorem (\ref{timeCompEstab}). This theorem 
establishes that, within an algorithmic descriptive complexity
 framework, similarly complex initial states must evolve into
 similarly complex future states over similarly complex time
frames, effectively erasing the difference between the complexity 
of the state of the system and the complexity of the corresponding 
time, and establishing absolute limits to the reducibility of 
future states.

Let $F(t)=T(S(M_0,E,t))$ be the \emph{real execution time} of the 
system at time $t$.  Using our time counting machine $U^H$, it is 
easy to see that $F(t)$ is computable and, given the uniqueness of 
the successor state, $F$ increases strictly with $t$, and hence is
 injective. Consequently, $F$ has a computational inverse $F^{-1}$ 
over its image. Therefore, we have it that (up to a small constant)
$K(F(t)) \leq K(F) + K(t)$ and $K(t) \leq K(F^{-1}) + K(F(t))$. 
It follows that $K(t) = K(F(t)) + O(c)$, where $c$ is an integer
 independent of $t$ (but that can depend on $S$). In other words, 
for a fixed system $S$, the execution time and the system time are
\emph{equally complex up to a constant}. From here on we will not 
differentiate between the complexity of both times. A 
generalization of the previous equation is given by the following 
theorem:

\begin{thm}[Time Complexity Stability]\label{timeCompEstab}
Let $S$ and $S'$ be two computable systems and $t$ and $t'$ the 
first time where each system reaches the states $M_t$ and
$M'_{t'}$ respectively. Then there exists $c$ such that $|K(M_t) -
K(t)|\leq c$ and $|K(M_t) - K(M'_{t'})|\leq c$. Specifically:

\begin{itemize}

\item[$i$)] There exists a natural number $c$ that depends on $S$
and $M_0$, but not on $t$, such that
\begin{equation}
|K(M_t) - K(t)|\leq c.
\end{equation}

\item[$ii$)] If $K(S(M_0,E,t))=K(S'(M'_0,E',t'))+O(1)$ and 
$K(M_0)=K(M'_0)+O(1)$ then there exists a constant $c$ that does 
not depend on $t$ such that $|K(t)-K(t')|\leq c$, where $t$ and
$t'$ are the minimum times for which the corresponding state is
 reached.

\item[$iii$)] Let $S$ and $S'$ be two dynamical systems with an 
infinite number of equally--up to a constant--descriptive complex 
times $\alpha_i$ and $\delta_i$. For any infinite subsequence of 
times with strictly growing descriptive complexity, all but 
finitely many $j,k$ such that $j>k$ comply with the equation: 
$K(\alpha_k)-K(\alpha_j) = K(\delta_k)-K(\delta_j)$.
\end{itemize}

\begin{proof}
First, note that we can describe a program such that given $S$,
$M_0$ and $E$, runs $S(M_0,E,x)$ for each $x$ until it finds $t$. Therefore
\begin{align}\label{eqTimeLimit}
K(t) \leq   K(S(M_0,E,t)) + K(S) + K(M_0) + K(E) + O(1),
\end{align}
\noindent{}.Similarly for $t'$. By the inequality \ref{timeDomIneq}
and the hypothesized equalities we obtain
\begin{align*}
K(t)  - ( K(S) + K(M_0) + K(E) + O(1)) \leq K(M_t) \leq K(t) + (K(S) + K(E) + K(M_0) + O(1)),
\end{align*}
\noindent{}which implies the first part. The second part is a 
direct consequence.

For the third part, suppose that there exists an infinity of times 
such that $K(\alpha_k)-K(\alpha_j)
> K(\delta_k) - K(\delta_j)$. Therefore $K(\alpha_k)- K(\delta_k)
> K(\alpha_j) - K(\delta_j)$, which implies that the difference is 
unbounded, which is a contradiction of the first part. Analogously, the 
other inequality yields the same contradiction.
\end{proof}
\end{thm}

The slow growth of time is a possible objection to the assertion that in the descriptive complexity 
of systems time is the dominating parameter for predicting their evolution:
the function $K(t)$ grows within an order of $O(\log t)$, which is
 very slow and often considered insignificant in the information theory 
literature. However, we have to consider the scale of time we are 
using. For instance, one second of \emph{real time} in the system 
we are modelling may mean an exponential number of discrete 
time steps for our computable model (for instance, if we are modelling a genetic 
machine with current computer technology), yielding a potential 
polynomial growth in their descriptive complexity. However, if 
this time conversion is computable, then $K(t)$ grows at most at a 
constant pace. This is an instance of \textit{irreducibility}, as 
there exist infinite sequences of times that cannot be obtained by computable 
methods. In the upcoming sections we will call such times \emph{random times} and the sequences containing them will be deemed \textbf{irreducible}.

\subsection{\label{nonRandom}Non-Randomness of Decidable Convergence Times}

One of the most important issues for science is predicting 
the future behaviour of dynamical systems. The prediction we 
will focus on is about the first state of convergence (definition
\ref{convergence}): Will a system converge and how long will it 
take? In this section we shall show the limit that 
decidability imposes on the complexity of the first convergent 
state. A consequences of this is the existence of undecidable 
adapted states.

 Formally, for the convergence of a system $S$ with degree
$\epsilon$ to be decidable there must exist an algorithm 
$D_\epsilon$ such that $D_\epsilon(S,M_0,E,\delta)=1$ if the 
system is convergent at time $\delta$ and $0$ otherwise. 
Moreover, we can describe a machine $P$ such that given full 
descriptions of $D_\epsilon$, $S$ and $M_0$ it runs $D_\epsilon$ 
with inputs $S$ and $M_0$ while running over all the possible 
times $t$, returning the first $t$ for which the system converges.
 Note that $\delta = P(\langle D_\epsilon \rangle \langle S \rangle
\langle M_0 \rangle \langle E \rangle)$. Hence we have a short
 description of $\delta$ and therefore $\delta$ \emph{cannot be
 random}: if $S(M_0,E,t)$ is a convergent system then
\begin{equation} \label{eqLimit}
K(\delta) \leq K(D_\epsilon) + K(S)  + K(E) + K(M_0) + O(1),
\end{equation}
where $\delta{}$ is the first time at which convergence is 
reached. Note that all the variables are known at the initial
 state of the system. This result can summed up by the following 
lemma:

\begin{lem}\label{lemEmg}
Let $S$ be a system convergent at time $\delta$. If $\delta$ 
is considerably more descriptively complex than the system and the 
environment, i.e. if for every \textit{reasonably large} natural 
number $d$ we have it that
\begin{equation*}
K(\delta) > K(S) + K(E) + K(M_0) + d,
\end{equation*}
\noindent{}then $\delta$ cannot be found by an algorithm described 
within $d$ number of characters.
\begin{proof}
It is a direct consequence of the inequality \ref{eqLimit}.
\end{proof}
\end{lem}
\noindent{}We call such times \emph{random convergence times} and 
the state of the system $M_\delta$ a \emph{random state}. It is 
important to note that the descriptive complexity of a random
 state must also be high:

\begin{lem}\label{lemCompState}
Let $S$ be a convergent system with a complex state
$S(M_0,E,\delta)$. For every \textit{reasonably large} $d$ we have it 
that
\begin{equation*}
K(S(M_0,E,\delta)) > K(S) + K(E) + K(M_0) + d.
\end{equation*}

\begin{proof}
Suppose the contrary to be true, i.e. that there exist $d$ small enough that
$K(S(M_0,E,\delta)) \leq K(S) + K(E) + K(M_0) + d$. Let $q$ be the 
program that, given $S$, $E$, $M_0$ and $S(M_0,E,\delta)$, runs
$S(M_0,E,t)$ in order for each $t$ and compares the result to 
$S(M_0,E,\delta)$, returning the first time where the equality is 
reached. Therefore, given the uniqueness of the successor state
(\ref{uSucP}), $\delta=q(S,M_0,E,S(M_0,E,\delta))$ and
\begin{align*}
K(\delta)  \leq & K(S) + K(E) + K(M_0) + K(S(M_0,E,\delta)) + |q|\\
  \leq & K(S) + K(E) + K(M_0) + (K(S) + K(E) + K(M_0) + d) + O(1),
\end{align*}
\noindent{}which gives us a small upper bound to the random 
convergence time $\delta$.
\end{proof}
\end{lem}

In other words, if $\delta$ has high descriptive complexity, then
there does not exist a reasonable algorithm that finds it even if 
we have a complete description of the system and its environment.
 It follows that the descriptive complexity of a computable
 convergent state cannot be much greater than the descriptive
 complexity of the system itself. 

What a \emph{reasonably large $d$} is has been handled so far with 
ambiguity, as it represents the descriptive complexity of any
 computable method $D_\epsilon$. We may intend to find convergence
 times, which intuitively cannot be arbitrarily large. It is easy
 to `\emph{cheat}' on the inequality \ref{eqLimit} by including in 
the description of the program $D_\epsilon$ the full description 
of the convergence time $\delta$, which is why we ask for
\emph{reasonable} descriptions. 

Another question left to be answered is whether complex convergence times do 
exist for a given limit $d$, considering that the limits imposed by the 
inequality \ref{eqLimit} loosen up in direct relation to the 
descriptive complexity of $S$, $E$ and $M_0$.

The next result answers both questions by proving the existence of
 complex convergence times for a broad characterization of the size
of $d$:

\begin{lem}[Existence of Random Convergence Times]\label{lemExist}
Let $F$ be a total computable function. For any $\epsilon$ there 
exists a system $S(M_0,E,t)$ such that the convergence times are
$F(S,M_0,E)$-random.

\begin{proof}
Let $E$ and $s$ be two natural numbers such that $K(E|s)> \epsilon$. 
By reduction to the Halting Problem (\cite{Turing}) it is easy to see 
the existence of $F(S,M_0,E)$-random convergence times: Let $T'$ be 
a Turing Machine, and $S_{t}$ the Turing machine that emulates $T$ 
for $t$ steps with input $M_0$ and returns $E$ for every time 
equal to or greater than the halting time, and $s$ otherwise. Let us 
consider the system $S(M_0,E,t)= S_{t}(\langle T \rangle \langle
M_0 \rangle \langle t \rangle \langle E \rangle )$. 

If the convergence times are not $F(S,M_0,E)$-random, then there 
exists a constant $c$ such that we can decide $HP$ by running $S'$
 for each $t$ that meets the inequality $|t| + 2\log |t| + c \leq |S'| + |\langle T \rangle \langle M_0
\rangle \langle t \rangle \langle E \rangle| +
F(S,M_0,E)\footnote{For any string $s$ there exists a self-delimited program (by a
`print') that takes a prefix-free input of the form
 $1^{\log |s|}0|s|s$.}$,
\noindent{}which cannot be done, since $HP$ is undecidable.
\end{proof}

\end{lem}

Let us focus on what the previous lemma is saying: $F$ can be any 
computable function. It can be a polynomial or exponential 
function with respect to the length of a given description for 
$M_0$ and $E$. It can also be any computable theory that we might 
propose for setting an upper limit to the size of an algorithm 
that finds convergence times given descriptions of the system's 
behaviour, environment and initial state. In other words, for a 
class of dynamical systems, finding convergence times, therefore 
convergent states, is not decidable, even with complete information
 about the system and its initial state. Finally, by the proof of the
 lemma, adapted states can be seen as a \textbf{generalization of 
halting states}.

\subsection{Randomness of Convergence in Dynamic 
Environments}\label{randDyn}

So far we have limited the discussion to fixed environments. 
However, as observed in the physical world, the environment itself 
can change over time. We call such environments \emph{dynamic
environments}. In this section we extend the previous results to 
cover environments that change depending on time as well as on the 
initial state of the system. We also propose a weaker convergence 
condition called \emph{weak convergence} and propose a necessary
(but not sufficient) condition for the computability of convergence 
times called \emph{descriptive differentiability}.

We can think of an environment $E$ as a dynamic computable system,
a moving target that also changes with time and depends on the 
initial state $M_0$. In order for the system to be convergent, we 
propose the same criterion---there must exist $\delta$ such that $n
\geq \delta$ implies
\begin{equation}
K(E(M_0,n)|S(M_0, E(M_0,n),n)) \leq \epsilon.
\end{equation}
\noindent{}A system with a dynamic environment also meets the 
inequality \ref{eqLimit} and lemmas \ref{lemEmg} and
\ref{lemExist} since we can describe a machine that runs both $S$
and $E$ for the same time $t$. Given that $E$ is a moving target it is convenient to consider an \textit{adaptation period} for the new states of $E$:

\begin{defn}\label{weakConvDef}
We say that $S$ \textit{converges weakly} to $E$ if there exist an 
infinity of times $\delta_i$ such that
\begin{equation}
K(E(M_0,{\delta_i})|S(M_0, E(M_0,{\delta_i}),\delta_i)) \leq
\epsilon.
\end{equation}
\end{defn}

As a direct consequence of the inequality \ref{eqLimit} and lemma
\ref{lemExist} we have the following lemma:

\begin{lem}\label{weakConvLemma}
Let $S(M_0, E(M_0,t),t)$ be a weakly converging system. Any
 decision algorithm $D_\epsilon(S, M_0, E, \delta_i)$ can only 
decide the first non-random time.
\end{lem}

As noted above, these results do not change when 
dynamic environments are considered. In fact, we can think of static 
environments as a special case of dynamic environments. However, 
with different targets of adaptability and convergence, it makes
 sense to generalize beyond the first convergence time. 
Also, it should be noted that specifying a convergence index adds 
additional information that a decision algorithm can potentially 
use.

\begin{lem}\label{weakConvLemma2}
Let $S(M_0, E(M_0,t),t)$ be a weakly converging system with an 
infinity of random times such that $k>j$ implies that $K(\delta_k)
= K(\delta_j) + \Delta K_\delta (j,k)$, where $\Delta K_\delta$ is
 a (not necessarily computable) function with a range confined to the positive 
integers. If the function $ \Delta K_\delta (i,i+m)$ is unbounded 
with respect to $i$, then any decision algorithm $D_\epsilon(S,
M_0, E, i)$, where $i$ is the $i$-th convergence time, can only
 decide a finite number of $i$s.

\begin{proof}
Suppose that $D_\epsilon(S, M_0, E, i)$ can decide an infinite 
number of instances. Let us consider two times $\delta_i$ and
$\delta_{i+m}$. Note that we can describe a program that, 
by using $D_\epsilon$, $S$, $E$ and $M_0$ and $i$ 
together with the distance $m$, finds $\delta_{i+m}$. The next 
inequality follows:
$$K(\delta_{i+m}) \leq K(D_\epsilon) + K(i) + K(m) +O(1).$$ 
Next, note that we can describe another program 
that given $\delta_i$ and using $D_\epsilon$, $S$, $E$ and $M_0$ 
finds $i$, from which 
$$K(i) \leq K(D_\epsilon) + K(\delta_i) +
O(1) \mbox{ and } - K(\delta_i) \leq K(D_\epsilon) - K(i) + O(1).$$
 Therefore:
$$\Delta K_\delta (i,i+m) = K(\delta_{i+m}) - K(\delta_i)
\leq 2K(D_\epsilon) + K(m) + O(1)$$
and $\Delta K_\delta (i,i+m)$ is bounded with respect to $i$.
\end{proof}

\end{lem}

\noindent{}We will say that a sequence of times $\delta_1,..,\delta_i,...$ is \textit{non-descriptively differentiable} if $\Delta K_\delta$ is not a total function, which, as a consequence of the previous lemma, implies non-computability of the sequence.

\begin{defn}\label{SeqTimeRand}
We say that a sequence of times 
$\delta_1,\delta_2,...,\delta_i,...$ is \emph{non-descriptively 
differentiable} if $\Delta K_\delta (m)$ is not a total function.
\end{defn}

\section{Beyond Halting States: Open-Ended Evolution}\label{OOE}

Inequality \ref{eqLimit} states that being able to predict or 
recognize adaptation imposes a limit to the descriptive complexity 
of the first adapted state. A particular case is the halting 
state, as shown in the proof of lemma \ref{lemExist}. In 
this section we extend the lemma to continuously evolving systems, 
showing that computability of adapted times limits the complexity 
of adapted states beyond the first, imposing a limit to open-ended
 evolution for three complexity measures: sophistication, coarse
 sophistication and busy beaver logical depth.

For a system in constant evolution converging to a dynamic 
environment, the lemma \ref{weakConvLemma2} imposes a limit to the 
growth of the descriptive complexity of a system with computable 
adapted states: \emph{if the growth of the descriptive complexity 
of a sequence of convergent times is unbounded in the sense of 
definition \ref{SeqTimeRand}, then all but a finite number of times 
are undecidable.} The converse would be convenient, however it is 
not always true. Moreover, the next series of results shows that 
imposing such a limit would impede strong OEE:

\begin{thm}\label{teoPInC}
Let $S$ be a non-cyclical computable system with initial state
$M_0$, $E$ a dynamic environment, and $\delta_1,...,\delta_i,...$ a 
sequence of times such that for each $\delta_i$ there exists a 
total function $p_i$ such that $p_i(M_{\delta_i})=E(\delta_i)$. If 
the function $p:i \mapsto p_i$ is computable, then the function
$\delta:i \mapsto \delta_i$ is computable.
\begin{proof}
Assume that $p$ is computable. We can describe a program 
$D_{\epsilon}$ such that, given $S$, $M_0$, $\delta_i$ and $E$, runs $p_{\delta_i}(M_t)$ and $E(t)$
 for each time $t$, returning 
$1$ if $\delta_i$-th $t$ is such that $p_{\delta_i}(t)=E(t)$, and
 $0$ otherwise. Therefore the sequence of $\delta_i$'s is 
computable.
\end{proof}
\end{thm}

The last result can be applied naturally to weakly convergent 
systems (\ref{weakConvDef}): the way each adapted
 state approaches to $E$ is unpredictable, in other words, its
\emph{behaviour} changes over different stages unpredictably. Formally:

\begin{cor}\label{teoPInCCor}
Let $S(M_0,E,t)$ be a weakly converging system, with adapted states
$M_{\delta_1},...,M_{\delta_i},...$ and $p_1,...,p_i,...$ its 
respective adapted behaviour. If the mapping $\delta:i \mapsto
\delta_i$ is non-descriptively differentiable then the function
$p:i \mapsto p_i$ is not computable.
\begin{proof}
It is a direct consequence of applying the theorem \ref{teoPInC} to 
the definition of weakly converging systems.
\end{proof}
\end{cor}

While asking for totality might look like an arbitrary limitation 
at first glance, the reader should recall that in weakly
 convergent systems the program $p_i$ represents an organism, a 
theory or other computable system that uses $M_{\delta_i}$'s
information to predict the behaviour of $E(\delta_i)$, and if this 
prediction does not process its environment in a sensible time
frame then it is hard to argue that it represents an \emph{adapted
system} or a \emph{useful theory}.

The intuition behind classifying descriptively differentiable 
adapted time sequences as \emph{less complex} is better explained
by borrowing ideas developed by Bennett and Koppel, within the 
framework of logical depth \cite{LDepth} and sophistication
\cite{koppelSoph}, respectively. Their argument states that
 random strings are as simple as very regular strings, given that 
there is no complex underlying structure in their minimal 
descriptions. The intuition that random objects contain no useful 
information leads us to the same conclusion. And given the theorem
\ref{timeCompEstab}, the states must retain a high degree of 
randomness for random times.

\emph{Sophistication} is a measure of \emph{useful information} 
within a string. Proposed by Koppel, the underlying approach consists in 
dividing the description of a string $x$ into two parts: the program 
that represents the \emph{underlying structure} of the object, and 
the input, which is the random or \emph{structureless} component 
of the object. This function is denoted by $soph_c(x)$, where $c$
is a natural number representing the significance level.

\begin{defn}\label{defSoph}
The \emph{sophistication} of a natural number $x$ at the 
significance level $c$, $c\in\mathbb{N}$, is defined as:
$$soph_c(x) = \min \{|\langle p \rangle | : \text{p is a total 
function and } \exists y.p(y)= x \text{ and } |\langle p
\rangle | + |y| \leq K(x) + c \}$$
\end{defn}

Now, the images of a mapping $\delta:i \mapsto \delta_i$ already 
have the form $\delta(i)$, where $\delta$ and $i$ represent the 
structure and the random component respectively. Random strings 
should bind this structure strongly up to a logarithmic error,
 which is proven in the next lemma.

\begin{lem}\label{lemSoph1}
Let $\delta_1,...,\delta_i,...$ be a sequence of different natural
numbers and $r$ a natural number. If the function $\delta : i
\mapsto \delta_i$ is computable then there exists an infinite
 subsequence where the sophistication is bounded up to an a
logarithm of a logarithmic term of their indexes.

\begin{proof}
Let $\delta$ be a computable function. Note that since $\delta$
is computable and the sequence is composed of different naturals, 
its inverse function $\delta^{-1}$ can be computed by a program 
$m$ which, given a description of $\delta$ and $\delta_i$,
 finds the first $i$ that produces $\delta_i$ and returns it; 
therefore $K(i) \leq K(\delta_i) + |\langle m \rangle | + |\langle
\delta \rangle |$ and $K(\delta) + K(i) \leq K(\delta_i) +
|\langle m \rangle | + 2|\langle \delta \rangle |$. Now, if $i$ is 
a $r$-random natural where the inequality holds tightly, we have 
it that
$(K(\delta)+ O(\log |i|)) + |i| - r \leq K(\delta_i) + |\langle m
\rangle | + 2|\langle \delta \rangle |$, which implies that, 
since $\delta$ is a total function, 
$soph_{(|\langle m \rangle | + 2|\langle \delta \rangle | +
r)}(\delta_i) \leq K(\delta) + O(\log \log i)$. Therefore, the 
sophistication is bounded up to an alogarithm of a logarithmic term
 for a constant significance level for an infinite subsequence.
\end{proof}

\end{lem}

Small changes in the significance level of sophistication can have
a large impact on the sophistication of a given string. Another 
possible issue is that the constant proposed in lemma
\ref{lemSoph1} could appear to be large at first (but it becomes 
comparatively smaller as $i$ grows). A \emph{robust} variation of 
sophistication called coarse sophistication \cite{antunesSop} incorporates the significance level as a penalty. The definition 
presented here differs slightly from theirs in order to maintain
 congruence with the chosen prefix-free universal machine and to 
avoid negative values. This measure is denoted by $csoph(x)$.

\begin{defn}\label{defCoarseSoph}
The \emph{coarse sophistication} of a natural number $x$ is 
defined as: $$ csoph(x) = \min \{ 2|\langle p
\rangle | + |\langle y \rangle | - K(x):p(y)=x \text{ and $p$
is total} \},$$ where $|\langle y \rangle |$ is a
computable unambiguous codification of $y$.
\end{defn}

With a similar argument as the one used to prove lemma
\ref{lemSoph1}, it is easy to show that coarse sophistication is 
similarly bounded up to an algorithm of a logarithmic term.

\begin{lem}\label{coarseSophLem}
Let $\delta_1,...,\delta_i,...$ be a sequence of different natural 
numbers and $r$ a natural number. If the function $\delta : i
\mapsto \delta_i$ is computable, then there exists an infinite
 subsequence where the coarse sophistication is bounded up to an a
lgorithm of a logarithmic term.

\begin{proof}
If $\delta$ is computable and $i$ is $r$-random, then by 
definition of $csoph$ and the inequalities presented in the proof 
of lemma \ref{lemSoph1}, we have it that
\begin{align*}
csoph(\delta_i)  \leq & 2K(\delta) + (|i| + 2 \log |i| +1 ) - K(\delta_i) \\
 \leq & 2K(\delta) + (|i| + 2 \log |i| +1 ) - K(i)  + |\langle M \rangle| + |\langle \delta \rangle| \\
 \leq & 2K(\delta) + |\langle M \rangle| + |\langle \delta \rangle|+ (|i| + 2 \log |i| +1
 )  - |i| + r\\
 = & 2K(\delta) + |\langle M \rangle| + |\langle \delta \rangle| + r + 1 + O( \log \log i))
\end{align*}
\end{proof}

\end{lem}

Another proposed measure of complexity is Bennett's logical depth
\cite{LDepth}, which measures the minimum computational time 
required to compute an object from a nearly minimal description. 
Logical depth works under the assumption that complex or
\emph{deep} natural numbers take a long time to compute from near 
minimal descriptions. Conversely, random or incompressible strings
are shallow since their minimal descriptions must contain the full 
description \emph{verbatim}. For the next result we will use a
related measure called busy beaver logical depth, denoted by
$depth_{bb}(x)$.

\begin{defn}\label{lDDef}
The \emph{busy beaver logical depth} of the description of a
natural $x$, denoted by $depth_{bb}(x)$, is defined as:
\begin{equation*}
depth_{bb}(x) = \min \{|p|-K(x)+j : U(p) = x \text{ and
}T(p)\leq BB(j)\},
\end{equation*}
\noindent{}where $T(P)$ is the halting time of the program $p$ and
$BB(j)$, known as the busy beaver function, is the halting time of
the slowest program that can be described within $j$ bits
\cite{dal82}.
\end{defn}

The next result follows from a theorem formulated by Antunes and
 Fortnow \cite{antunesSop} and from lemma \ref{coarseSophLem}.
\begin{cor}\label{LogicalDepLem}
Let $\delta_1,...,\delta_i,...$ be a sequence of different natural 
numbers and $r$ a natural number. If the function $\delta : i
\mapsto \delta_i$ is computable, then there exists an infinite
 subsequence where the busy beaver logical depth is bounded up to an algorithm of a logarithmic term of their indexes.

\begin{proof}
By theorem 5.2 at \cite{antunesSop}, for any $i$ we have it that
$|csoph(\delta_i)-depth_{bb}(\delta_i)| \leq O(\log |\delta_i|).$ By lemma
\ref{coarseSophLem} and theorem \ref{timeCompEstab} the result 
follows.
\end{proof}
\end{cor}

Let us focus on the consequence of lemmas \ref{lemSoph1} and
\ref{coarseSophLem} and corollary \ref{LogicalDepLem}. Given the 
relationship established between descriptive time complexity and
the corresponding state of a system (theorem \ref{timeCompEstab}),
these last results imply that either the complexity of the adapted
 states of a system (using any of the three complexity measures)
grows very slowly for an infinite subsequence of times (becoming
increasingly common up to a probability limit of 1 \cite{stopQ}) or
 the subsequence of adapted times is undecidable.

\begin{thm}\label{teoOEELim}
If $S(M_0,E(t),t)$ is a weakly converging system with adaptation
 times $\delta_1,...,\delta_i,...$ that exhibits strong OEE with respect to $csoph$ and $depth_{bb}$, 
then the mapping $\delta: i \mapsto \delta_i$ is not computable. Also, there exists a constant $c$ such that the result applies to $soph_c$.
\begin{proof}
We can see the sequence of adapted states as a function
$M_{\delta_i}:i \mapsto M_{\delta_i}$. By lemmas \ref{lemSoph1}
and \ref{coarseSophLem} and corollary \ref{LogicalDepLem}, for the
 three stated measures of complexity, there exists an infinite
subsequence where the respective complexity is upper bounded by
$O(\log \log i)$. It follows that if the complexity grows faster 
than $O(\log \log i)$ for an infinite subsequence, then there must
 exist an infinity of indexes $j$ in the bounded succession where 
$\gamma(j)$ grows faster than $C(M_j)$. Therefore there exists an
 infinity of indexes $j$ where $C(M_j) - \gamma(j)$ is upper
bounded. Finally, note that if a computable mapping $\delta: i \mapsto \delta_i$ allows growth on the order of $O(\log \log i)$, then the computable function $\delta': i \mapsto \delta_{2^{2^i}}$ would grow faster than the stated bound.
\end{proof}
\end{thm}

Now, in the absence of absolute solutions to the
 problem of finding adapted states in the presence of strong OEE,
 one might cast about for a partial solution or approximation that
 decides most (or at least some) of the adapted states. The
 following corollary shows that the problem is not even
semi-computable: \emph{any algorithm one might propose can only
decide a bounded number of adapted states}.

\begin{cor}\label{semiCompLimit}
If $S(M_0,E,t)$ is a weakly converging system with adapted states 
$M_1,...,M_i,...$ that show strong OEE, then the mapping $\delta: i \mapsto \delta_i$ 
is not even semi-computable.

\begin{proof}
Note that for any subsequence of adaptation times
$\delta_{j_1},...,\delta_{j_k},...$, the system must show strong 
$OEE$. Therefore, by theorem \ref{teoOEELim}, any subsequence must 
also not be computable. It follows that there cannot exist an 
algorithm that produces an infinity of elements of the sequence,
since such an algorithm would allow the creation of a computable
 subsequence of adaptation times.
\end{proof}
\end{cor}

In short, the theorem \ref{teoOEELim} imposes undecidability
 on strong OEE and, according to theorem \ref{teoPInCCor}, the behaviour and 
interpretation of the system evolves in an unpredictable way, 
establishing one path for \emph{emergence}: \emph{a set of rules
 for future states that cannot be reduced to an initial set of 
rules}. Recall that for a given weakly converging dynamical 
system, the sequence of programs $p_i$ represents the behaviour or
interpretation of each adapted state $M_i$. If a system exhibits 
strong OEE with respect to the complexity measures $soph_c$,
 $csoph$ or $depth_{bb}$, by corollary \ref{teoPInCCor} and theorem
\ref{teoOEELim} the sequence of behaviours is uncomputable, and 
therefore irreducible to any function of the form $p:i \mapsto
p_i$, even when possessing complete descriptions for the behaviour 
of the system, its environment and its initial state. In other
 words, \emph{the behaviour of iterative adapted states cannot be
obtained from the initial set of rules}. Furthermore, we conjecture 
that the results hold for all adequate measures of complexity:

\begin{conj}\label{con:conj}
Computability bounds the growing complexity rate to that of an
order of the slowest growing infinite subsequence with respect to
any \emph{adequate} complexity measure $C$.
\end{conj}

\subsection{A System Exhibiting OEE}

With the aim of providing mathematical evidence for the adequacy
of Darwinian evolution, Chaitin developed a mathematical model 
that converges to its environment significantly faster than 
exhaustive search, being fairly close to an \emph{intelligent}
solution to a mathematical problem that requires \emph{maximal
creativity} \cite{chaitin:EvolofMutaSoft,ChaitinBook}. 

One of the solutions Chaitin proposes is to find digital organisms that 
approximate the busy beaver function:
$$BB(n) = \max\{T(U(p)): |p|\leq n\},$$ which is equivalent
(up to a constant) to asking for the largest natural number 
that can be named within $n$ number of bits and the first $n$ bits of 
Chaitin's constant, which is defined as $\Omega_U=\scriptstyle \sum_{T\in{}HP} 2^{-|T|}$, where $HP$ is the set of all halting Turing machines for the universal machine $U$. We will omit the subindex from $\Omega$ in the rest of this text. 

Chaitin's evolutionary system searches non-deterministically 
through the space of Turing machines using a reference universal
 machine $U'$ with the property that all strings are valid 
programs. This random walk starts with the empty string 
$M_0=\text{``''}$, and each new state is defined 
as the output of a Turing machine, called a \textit{mutation}, with the previous state 
as an input. These mutations are chosen stochastically according to the universal distribution \cite{kirchherr1997miraculous}. If these mutations help to more accurately approximate the digits of $\Omega$,
 then this program becomes the new state $M_{t+1}$, otherwise we keep searching for new organisms. Chaitin demonstrates that the system approaches
$\Omega$ \textit{efficiently} (with quadratic overhead),
arguing that this is evidence of the adequacy of Darwinian evolution \cite{ChaitinEvolvingSoftware}.

Given that $\Omega$ can be used to compute $BB(n)$  \cite{Gardner79}, a deterministic version of Chaitin's system is the following:
\begin{align*}
M_0 & =  0 \\
M_t & = p.T(p)= \max \{T(U'(q)):H(M_{t-1},q)\leq w\},
\end{align*}
where $H(M_{t-1},p)$ is the \textit{distance} between the programs
$M_{t-1}$, $q$ is the quantification of the number of mutations needed to 
transform one string into the other, and $w$ is a positive integer 
acting as an accumulator that resets to 1 whenever $M_t$ increases 
in value, adding 1 otherwise. 

Defining a computable environment or adaptation condition for this 
system is difficult since the system seeks to approach an 
uncomputable function ($BB$) and the evolution rule itself is not 
computable given the halting problem. The most direct way to 
define it is $E(t)=BB(t)$ or, equivalently, as the first $t$-bits
of Chaitin's constant $\Omega$.

Another way to define the environment is by an encoding of the proposition \textit{larger than $U(M_{t-1})$} for each time $t$. Given that we can compute $M_{t-1}$ and its relationship with $M_t$ given a description of the latter and a constant amount of information ($\epsilon$), we find adaptation at the times $t$ where the busy beaver function grows.

It is easy to see that the sequence of programs $i \mapsto M_i$ is 
precisely what generates the busy beaver sequence 
$\eta_i=BB(i)$. Given that $BB(t)$ is not a computable function,
the evolution of the system, along with the respective adaptation
 times, is not computable. Furthermore, this sequence is composed
of programs that compute, in order, an element of a sequence that exhibits strong OEE with respect to $depth_{bb}$: let $\eta_i=BB(i)$ be the sequence of 
all busy beaver values; by definition, if $i$ is the first 
value for which $BB(i)$ was obtained,
$depth_{bb}(BB(i))  = \min\{|p_i| - K(BB(i)) + i\}$,
where $U(P_i)=K(BB(i))$. It follows that  $K(BB(i))=|p_i|$ and 
$depth_{bb}(BB(i))=i$, otherwise $p_i$ would not be the minimal
 program.

Computing the system described requires a solution for the Halting Problem, and the system itself might also seem \textit{unnatural} at first glance. However, we can think of the biosphere as a huge parallel computer that is constantly approximating solutions to the adaptation problem by means of survivability, and just as $\Omega$  has been approximated \cite{calude2002computing}, we claim that \textit{just as we cannot know whether a Turing machine will halt until it does, we may not know if an organism will keep adapting and survive in the future, but we can know when it failed to do so (extinction)}.
  
\section{Logical Depth and Future Work}

Although we conjecture that the theorem \ref{teoOEELim} must also
hold for logical depth as defined by Bennett \cite{LDepth}, 
extending the results to this measure is still a work in progress.
Encompassing logical depth will require a deeper understanding of
 the internal structure of the relationship between system and 
computing time, beyond the time complexity stability
(\ref{timeCompEstab}), and might be related to open fundamental 
problems in computer science and mathematics. For instance,
finding a \emph{low} upper bound to the growth of logical depth of
all computable series of natural numbers would suggest a negative
answer to the question of the existence of an efficient way of
generating deep strings, which Bennett relates to the $P \neq
PSPACE$ problem.

 One way to understand conjecture \ref{con:conj} is that \emph{the 
information of future states of a system is either contained at 
the initial state--hence their complexity is bounded by that 
initial state-- or is undecidable}. This should be a consequence 
given that, for any computable dynamical system, the randomness
induced by time cannot be avoided.

Given that we intend to expand upon these questions in the future, 
it is important to address the fact that the diagonal algorithm that 
Bennett proposes for generating deep strings represents a contradiction 
to our conjecture: The \emph{logical depth} of a natural $x$ at
the level of significance $c$ is defined as:
\begin{equation*}
\begin{multlined}[t]
depth_{c}(x) = \min \{T(p):|p|-K(x)<c \text{ and } U(p) =
x\}.
\end{multlined}
\end{equation*}
The algorithm $\chi(n,T)$ produces strings of length $n$ with
depth $T$ for a significance level $n-K(T)-O(\log n)$, where
$K(T)$ must be smaller than $n$, and $n$ must not be \emph{as large}
(or larger) than  $T$ to avoid shallow strings. One possible issue 
with this algorithm is that the significance level is not
 computable, and we can expect it to vary greatly with respect to
$K(T)$: For large $T$ with small $K(T)$ (such as $T^{T^T}$) the 
significance level is nearly $n$, which suggests that, for a
\emph{steady} significance level with respect to times $T$ with 
large $K(T)$, the growth in complexity might not be stable. This 
issue, along with an algorithm that consistently enumerates pairs 
of $n$ and $T$s such that $K(T)<n<<T$ for growing $T$'s, will be 
explored in future work and its solution would require a formal 
definition of \emph{adequate} complexity measures. The fact that $\chi$ 
presents a challenge to the conjecture \ref{con:conj} would suggest an important difference from the
 three complexity measures used in this article.

\section{Conclusions}

We have presented a formal and general
mathematical model for adaptation within the framework of
computable dynamical systems. This model exhibits universal
properties for all computable dynamical systems, of which Turing machines are a subset. Among other results, we have given formal definitions of 
open-ended evolution (OEE) and strong open-ended evolution and supported the latter on the basis that it allows us to differentiate between trivial and non-trivial systems. 

We have also shown that decidability imposes universal limits on the growth of complexity in computable systems, as measured by sophistication, coarse sophistication and busy beaver logical depth. We show that as time dominates the descriptive algorithmic complexity of the states, the complexity of the evolution of a system tightly follows that of natural numbers, implying the existence of non-trivial states but the non-existence of an algorithm for finding these states or any subsequence of them, which makes the computations for harnessing or identifying them undecidable.

Furthermore, as a direct implication of corollary
\ref{teoPInCCor} and theorem \ref{teoOEELim}, the undecidability of 
adapted states and the unpredictability of the behaviour of the system 
at each state is a requirement for a system to exhibit strong 
open-ended evolution with respect 
to the complexity measures known as sophistication, coarse
 sophistication and busy beaver logical depth, providing 
rigorous proof that undecidability and irreducibility of future 
behaviour is a requirement for the growth of complexity in the 
class of computable dynamical systems. We conjecture that these results can be extended to any adequate  complexity measure that assigns low complexity to random objects. Finally, we provide an example of a (non-computable) system that exhibits strong OEE and supply arguments for its adequacy as a model of evolution, which we claim supports our characterization of strong OEE.

\subsubsection*{Acknowledgements} We would like to thank Carlos
Gershenson Garc\'{i}a for his comments during the 
development of this project and to acknowledge support from grants
 CB-2013-01/221341 and PAPIIT IN113013.

\bibliographystyle{plain}
\bibliography{ref}

\begin{thebibliography}{10}

\bibitem{Adami94a}
C.~Adami and C.~T. Brown.
\newblock Evolutionary learning in the 2{D} artificial life system avida.
\newblock In {\em Proc. Artificial Life IV}, pages 377--381. MIT Press, 1994.

\bibitem{adami2002complexity}
Christoph Adami.
\newblock What is complexity?
\newblock {\em BioEssays}, 24(12):1085--1094, 2002.

\bibitem{alyssa}
A.~Adams, H.~Zenil, P.W.C. Davies, and S.I. Walker.
\newblock Formal definitions of unbounded evolution and innovation reveal
  universal mechanisms for open-ended evolution in dynamical systems.
\newblock {\em Scientific Reports (in press)}, 2016.

\bibitem{antunesSop}
L.~Antunes and L.~Fortnow.
\newblock Sophistication revisited.
\newblock In {\em ICALP: Annual International Colloquium on Automata, Languages
  and Programming}, 2003.

\bibitem{AuerbachB14}
Joshua~Evan Auerbach and Josh~C. Bongard.
\newblock Environmental influence on the evolution of morphological complexity
  in machines.
\newblock {\em PLoS Computational Biology}, 10(1), 2014.

\bibitem{Bedau98}
Bedau.
\newblock Four puzzles about life.
\newblock {\em ARTLIFE: Artificial Life}, 4, 1998.

\bibitem{BedauEtAl00}
Bedau, McCaskill, Packard, Rasmussen, Adami, Green, Ikegami, Kaneko, and Ray.
\newblock Open problems in artificial life.
\newblock {\em ARTLIFE: Artificial Life}, 6, 2000.

\bibitem{LDepth}
C.~H. Bennett.
\newblock Logical depth and physical complexity.
\newblock In R.~Herken, editor, {\em The Universal Turing Machine: A
  Half-Century Survey}, pages 227--257. Oxford University Press, 1988.

\bibitem{BBKT00}
Vincent~D. Blondel, Olivier Bournez, Pascal Koiran, and John~N. Tsitsiklis.
\newblock {The stability of saturated linear dynamical systems is undecidable}.
\newblock In Horst Reichel~Sophie Tison, editor, {\em {Symposium on Theoretical
  Aspects of Computer Science (STACS), Lille, France}}, volume 1770 of {\em
  Lecture Notes in Computer Science}, pages 479--490. Springer-Verlag, Feb
  2000.

\bibitem{CIE2013Daniel}
Olivier Bournez, Daniel~S. Gra{\c c}a, Amaury Pouly, and Ning Zhong.
\newblock Computability and computational complexity of the evolution of
  nonlinear dynamical systems.
\newblock In Springer, editor, {\em Computability in Europe (CIE'2013)},
  Lecture Notes in Computer Science, 2013.

\bibitem{Bournez}
Olivier Bournez, Daniel~S. Gra{\c{c}}a, Amaury Pouly, and Ning Zhong.
\newblock {\em The Nature of Computation. Logic, Algorithms, Applications: 9th
  Conference on Computability in Europe, CiE 2013, Milan, Italy, July 1-5,
  2013. Proceedings}, chapter Computability and Computational Complexity of the
  Evolution of Nonlinear Dynamical Systems, pages 12--21.
\newblock Springer Berlin Heidelberg, Berlin, Heidelberg, 2013.

\bibitem{calude2002computing}
Cristian~S Calude, Michael~J Dinneen, Chi-Kou Shu, et~al.
\newblock Computing a glimpse of randomness.
\newblock {\em Experimental Mathematics}, 11(3):361--370, 2002.

\bibitem{stopQ}
Cristian~S. Calude and Michael Stay.
\newblock Most programs stop quickly or never halt.
\newblock {\em CoRR}, abs/cs/0610153, 2006.

\bibitem{complexIncomplt}
C.S. Calude and H.~Jugensen.
\newblock Is complexity a source of incompleteness?
\newblock {\em ADVAM: Advances in Applied Mathematics}, 35, 2005.

\bibitem{Chaitin3}
G.~J. Chaitin.
\newblock Algorithmic information theory.
\newblock In {\em Encyclopaedia of Statistical Sciences}, volume~1, pages
  38--41. Wiley, 1982.

\bibitem{ChaitinEvolvingSoftware}
Gregory Chaitin.
\newblock Life as evolving software.
\newblock In Hector Zenil, editor, {\em A Computable Universe: Understanding
  and Exploring Nature as Computation}, chapter~16. World Scientific Publishing
  Company, 10 2012.

\bibitem{ChaitinBook}
Gregory Chaitin.
\newblock {\em Proving Darwin: Making Biology Mathematical}.
\newblock Vintage, 2013.

\bibitem{chaitin:EvolofMutaSoft}
Gregory~J. Chaitin.
\newblock Evolution of mutating software.
\newblock {\em Bulletin of the EATCS}, 97:157--164, 2009.

\bibitem{Cooper:Emergence}
S.~Barry Cooper.
\newblock Emergence as a computability-theoretic phenomenon.
\newblock {\em Applied Mathematics and Computation}, 215(4):1351--1360, 2009.

\bibitem{dal82}
R.~Daley.
\newblock Busy beaver sets: Characterizations and applications.
\newblock {\em INFCTRL: Information and Computation (formerly Information and
  Control)}, 52:52--67, 1982.

\bibitem{SymbDynSystm}
Jean-Charles Delvenne, Petr Kurka, and Vincent~D. Blondel.
\newblock Decidability and universality in symbolic dynamical systems.
\newblock {\em Fundam. Inform}, 74(4):463--490, 2006.

\bibitem{Fredkin-Toffoli-82}
E.~Fredkin and T.~Toffoli.
\newblock Conservative logic.
\newblock {\em International Journal of Theoretical Physics}, 21:219--253,
  1982.

\bibitem{Gardner79}
Gardner.
\newblock Mathematical games: The random number omega bids fair to hold the
  mysteries of the universe.
\newblock {\em SCIAM: Scientific American}, 241, 1979.

\bibitem{kirchherr1997miraculous}
Walter Kirchherr, Ming Li, and Paul Vit{\'a}nyi.
\newblock The miraculous universal distribution.
\newblock {\em The Mathematical Intelligencer}, 19(4):7--15, 1997.

\bibitem{Kolmogorov}
Andrey Kolmogorov.
\newblock Three approaches to the quantitative definition of information.
\newblock {\em Problems Inform. Transmission}, 1:1--7, 1965.

\bibitem{koppelSoph}
M.~Koppel.
\newblock Structure.
\newblock In R.~Herken, editor, {\em The Universal Turing Machine: A
  Half-Century Survey}, pages 435--452. Oxford University Press, 1988.

\bibitem{expltOEE}
Joel Lehman and Kenneth~O. Stanley.
\newblock Exploiting open-endedness to solve problems through the search for
  novelty.
\newblock In Seth Bullock, Jason Noble, Richard~A. Watson, and Mark~A. Bedau,
  editors, {\em ALIFE}, pages 329--336. MIT Press, 2008.

\bibitem{IntKolmogorov}
M.~Li and P.~Vit\'anyi.
\newblock {\em An introduction to {K}olmogorov complexity and its
  applications}.
\newblock Springer, 2nd edition, 1997.

\bibitem{lindgren:1992:epsd}
Kristian Lindgren.
\newblock Evolutionary phenomena in simple dynamics.
\newblock In Christopher~G. Langton, Charles Taylor, J.~Doyne Farmer, and Steen
  Rasmussen, editors, {\em Artificial Life {II}}, pages 295--312.
  Addison-Wesley, Redwood City, CA, 1992.

\bibitem{Margolus84}
N.~Margolus.
\newblock Physics-like models of computation.
\newblock {\em Physica D}, 10:81--95, 1984.
\newblock {\it ~\\ Discussion of reversible cellular automata illustrated by an
  implementation of Fredkin's Billiard-Ball model of computation.}

\bibitem{dynamicS}
J.~Meiss.
\newblock Dynamical systems.
\newblock {\em Scholarpedia}, 2(2):1629, 2007.
\newblock {revision \#121407}.

\bibitem{Moore}
Christopher Moore.
\newblock Generalized shifts: {U}npredictability and undecidability in
  dynamical systems.
\newblock {\em Nonlinearity}, 4(2):199--230, 1991.

\bibitem{Univdeflife}
K.~Ruiz-Mirazo, J.~Peret{\'o}, and A.~Moreno.
\newblock A universal definition of life: Autonomy and open-ended evolution.
\newblock {\em Origins of life and evolution of the biosphere}, 34(3):323--346,
  2002.

\bibitem{Soros:2014}
L.~B. Soros and Kenneth~O. Stanley.
\newblock Identifying necessary conditions for open-ended evolution through the
  artificial life world of chromaria.
\newblock In {\em Fourteenth International Conference on the Synthesis and
  Simulation of Living Systems (ALIFE 14)}. MIT Press, 2014.

\bibitem{openEndedArtEv}
Russell~K. Standish.
\newblock Open-ended artificial evolution.
\newblock {\em International Journal of Computational Intelligence and
  Applications}, 3(2):167--175, 2003.

\bibitem{TaylorOEEW}
Tim Taylor.
\newblock Requirements for open-ended evolution in natural and artificial
  systems.
\newblock {\em CoRR}, abs/1507.07403, 2015.

\bibitem{Turing}
A.~M. Turing.
\newblock On computable numbers, with an application to the
  entscheidungsproblem.
\newblock {\em Proceedings of the London Mathematical Society}, 42:230--265,
  1936.

\bibitem{zenil1}
H{\'e}ctor Zenil, Carlos Gershenson, James A.~R. Marshall, and David~A.
  Rosenblueth.
\newblock Life as thermodynamic evidence of algorithmic structure in natural
  environments.
\newblock {\em Entropy}, 14(11), 2012.

\end{thebibliography}

\end{document}